\documentstyle[aps]{revtex}

\input{psfig}

\begin{document}
\draft
\preprint{}

\def\Journal#1#2#3#4{{#1} {\bf #2}, #3 (#4)}

\def\NCA{\em Nuovo Cimento}
\def\NIM{\em Nucl. Instrum. Methods}
\def\NIMA{{\em Nucl. Instrum. Methods} A}
\def\NPB{{\em Nucl. Phys.} B}
\def\PLB{{\em Phys. Lett.}  B}
\def\PRL{\em Phys. Rev. Lett.}
\def\PRD{{\em Phys. Rev.} D}
\def\ZPC{{\em Z. Phys.} C}

\def\st{\scriptstyle}
\def\sst{\scriptscriptstyle}
\def\mco{\multicolumn}
\def\epp{\epsilon^{\prime}}
\def\vep{\varepsilon}
\def\ra{\rightarrow}
\def\ppg{\pi^+\pi^-\gamma}
\def\vp{{\bf p}}
\def\ko{K^0}
\def\kb{\bar{K^0}}
\def\al{\alpha}
\def\ab{\bar{\alpha}}
\def\be{\begin{equation}}
\def\ee{\end{equation}}
\def\bea{\begin{eqnarray}}
\def\eea{\end{eqnarray}}
\def\CPbar{\hbox{{\rm CP}\hskip-1.80em{/}}}


\title{HADRONIC AND ELEMENTARY MULTIPLICITY DISTRIBUTIONS IN A 
GEOMETRICAL APPROACH}

\author{P. C. BEGGIO, M. J. MENON}

\address{Instituto de F\'{\i}sica ``Gleb Wataghin'',\\
Universidade Estadual de Campinas, Unicamp,\\ 
13083-970 Campinas, SP, Brazil\\E-mail: menon@ifi.unicamp.br} 

\author{P. VALIN}

\address{D\'ept. de Physique, Universit\'e de Montr\'eal,\\
CP. 6128, Succ. 
Centre-Ville\\
Montr\'eal, Qu\'ebec, Canada H3C 3J7\\E-mail: pierre.valin@lmco.com}


\maketitle

\begin{abstract}
 We construct the hadronic multiplicity 
distribution in terms of an 
elementary distribution (at a given impact parameter) and the 
inelastic overlap function characterized by the observed BEL 
(Blacker-Edgier-Larger) behaviour. With suitable parametrizations 
for the elementary quantities, based on some geometrical arguments 
and the most recent data on $e^{+}e^{-}$ annihilation, an excellent 
description of $pp$ and $\overline{p}p$ inelastic multiplicity 
distributions at the highest energies is obtained.

\end{abstract}

\section{Introduction}

It is well known that the underlying theory of hadronic interactions, 
the quantum chromodynamics (QCD) is not yet able, by its own, to 
describe 
the bulk of experimental data associated with soft (long range) 
processess, 
in particular elastic scattering. At present, phenomenological 
approaches are very important as a source of information 
for adequate theoretical developments. On the other hand, from the 
experimental 
point of view, a renewed interest on both $pp$ and $\overline{p}p$ 
elastic 
and inelastic scattering is expected with the advent of the next 
accelerator 
generation, the RHIC and LHC \cite{acc}.
At this stage, due to our limited theoretical understanding of 
elastic 
scattering and, on the contrary, the success of QCD in treating 
hard (inelastic) 
processes, it may be important to investigate possible {\em 
connections between elastic and inelastic channels}, even from 
a phenomenological point of view. 
In this work we shall treat this subject in the contexts of the 
impact 
parameter picture, unitarity and the eikonal approximation. 

Following other 
authors \cite{bly}, we express the ``complex" (overall) 
hadron-$p$ multiplicity 
distributions (inelastic channel) in terms of an ``elementary" 
distribution 
associated with an elementary process taking place at given 
impact parameter and the inelastic overlap function (which is 
constructed from elastic 
channel data). The novel aspects concern: (a) quantitative 
correlation between 
the {\it violations} of the KNO scaling \cite{kno} (inelastic 
channel) and geometrical 
scaling \cite{jorge} (elastic channel); (b) introduction of 
suitable parametrizations 
for the elementary quantities, based on some geometrical arguments 
and the most 
recent data on $e^{+}e^{-}$ annihilation. With this formalism, the 
hadronic 
multiplicity distribution may be evaluated without any free 
parameter and an 
excellent reproduction of the experimental data on $pp$ (ISR) and 
$\overline{p}p$ ($S\overline{p}pS$) inelastic multiplicities is 
achieved.
A detailed discussion on all the results presented here and also a
complete list of references to the experimental data may be found in
a recent paper \cite{bmv}.

\section{Impact Parameter Formalism}

In the geometrical picture, unitarity correlates the elastic 
scattering 
amplitude in the the impact parameter $b$ space, $\Gamma(b,s)$, 
with the 
inelastic overlap function, $G_{in}(b,s)$, by
$2 Re \Gamma(b,s) = |\Gamma(b,s)|^2 + G_{in}(b,s)$.
For a purely imaginary elastic amplitude in momentum transfer 
space the 
profile function $\Gamma(b,s)$ is real and in the eikonal 
approximation 
is expressed by
$\Gamma(b,s)=1-exp[-\Omega(b,s)]$,
so that

\begin{equation} 
G_{in}(b,s)=1-exp[-2\ \Omega(b,s)] \equiv \sigma_{in}(b,s)
\end{equation}
is the probability for an inelastic event to take place at $b$ 
and $s$, namely, 
$\sigma_{in}(s)=\int d^{2}{\bf b}\ G_{in}(b,s)$.
In this picture the topological cross section for producting an 
even 
number $N$ of charged particles at $\sqrt{s}$ may be expressed by 
$
\sigma_{N}(s)= \int d^{2}{\bf b}\ \sigma_{N}(b,s) =  \int 
d^{2}{\bf b}\ 
\sigma_{in}(b,s) \sigma_{N}(b,s) / \sigma_{in}(b,s)
$. 

In this context, the formal connection between hadronic and 
elementary multiplicity distributions is obtained as follows.
Let $\varphi$ be the elementary multiplicity distribution, 
$<n>(b,s)$ the average number of particles produced at $b$ and 
$s$ and  
$z=N(s)/<n>(b,s)$
a KNO variable associated with the elementary process taking 
place at 
$b$ (and $s$). Then, in general, 
$ 
\varphi=<n>(b,s) \sigma_{N}(b,s) / \sigma_{in}(b,s) = \varphi 
(z,s).
$
Representing the overall multiplicity distribution by $\Phi$ and the 
corresponding KNO variable by  
$Z=N(s)/<N>(s)$,
where $<N>(s)$ is the average multiplicity at $\sqrt{s}$, we have 
in general 
$ 
\Phi=<N>(s) \sigma_{N}(s) / \sigma_{in}(s) = \Phi (Z,s)
$.
Both distributions are normalized by the usual conditions \cite{bly}

\begin{equation}
\int_{0}^{\infty} \varphi(z) dz=2=\int_{0}^{\infty} \varphi(z) z dz,
\qquad
\int_{0}^{\infty} \Phi(Z) dZ=2=\int_{0}^{\infty} \Phi(Z) Z dZ.
\end{equation}

Now, introducting a {\em multiplicity function} as the ratio

\begin{equation} 
{<n>(b,s) \over <N>(s)} \equiv m(b,s),
\end{equation}
the relationship between $\Phi$ and $\varphi$ follows from the above 
equations:

\begin{equation} 
\Phi=\frac{\int d^{2} {\bf b} \frac{G_{in}(b,s)}{m(b,s)} \varphi 
(\frac{Z}{m(b,s)})}{ \int d^{2} {\bf b}\  G_{in} (b,s)} = \Phi (Z,s).
\end{equation}

This result means that, once one has parametrizations for 
$G_{in}(b,s)$ and the elementary quantities $\varphi$ (multiplicity 
distribution) and $m(b,s)$ (multiplicity function) the overall 
hadronic
 multiplicity distribution may be evaluated. In this work we 
consider 
$G_{in}(b,s)$ from analyses of elastic $pp$ and $\overline{p}p$ 
scattering 
data (taking account of geometrical scaling violation) and infer 
the 
elementary quantities based on geometric arguments and experimental 
data 
on $e^{+}e^{-}$ annihilation, as explained in what follows. 

In the elastic channel, the breaking of Geometrical scaling is 
quite well 
described by the BEL behaviour, analytically expressed by the 
Short Range
 Expansion \cite{hv},
$G_{in}(b,s)=P(s)exp\{-b^{2}/4B(s)\}k(x,s)$,
with $k$ expanded in terms of a short-range variable $x=b\ 
exp\{-(\gamma b)^{2}/4 B(s)\}$. With suitable parametrizations 
for
$P(s)$ and $B(s)$ an excellent agreement with experimental data 
on
$pp$ and $\overline{p}p$ elastic scattering is achieved and we 
shall use
this well known result \cite{hv} as the input from the elastic 
channel.

\section{Elementary Hadronic Process}

We now turn to the discussion of the elementary hadronic 
processes, 
characterized by $\varphi$ and $m$ in Eq. (4).
At a given impact parameter, the most elementary process is 
an 
$e^{+}e^{-}$ collision, which is a process occuring in a 
unique 
angular momentum state and therefore also at a given impact 
parameter 
(zero in this case). Although $e^{+}e^{-}$ annihilations 
cannot be 
exactly the same as collisions between hadron constituents, 
it is 
reasonable, even from a geometric point of view, to think 
that some 
characteristics of both processes may be similar. The point 
is to find 
out or infer what they could be.

First, let us consider that the average multiplicity at given 
impact 
parameter depends on the center-of-mass energy in the form of 
a general 
power law 
$<n>(b\ fixed,s) \propto E_{CM}^{\gamma}$.
Now, from  Eq. (1), $exp\{-2\ 
\Omega(b,s)\}$ is the transmission coefficient, i.e. the 
probability of 
having no interaction at a given impact parameter, and therefore 
$\Omega$ 
should be proportional to the thickness of the target, or the 
energy 
$E_{CM}$ that can be deposited at $b$ for particle production 
at a given 
$s$. Then, we can express
$<n>(b,s) \propto \Omega^{\gamma}(b,s)$ and
comparison with Eq. (3), allows us to infer the multiplicity 
function: 

\begin{equation}
m(b,s)=\xi \Omega^{\gamma}(b,s),
\end{equation}
with $\xi$ being calculated by the normalization condition of the 
overall 
multiplicity distribution, Eq. (2). With this, Eq. (4) becomes

\begin{equation} 
\Phi=\frac{\int d^{2} {\bf b}\ \frac{G_{in}(b,s)}{\xi 
\Omega^{\gamma}(b,s)}\ 
\varphi (\frac{Z}{\xi \Omega^{\gamma}(b,s)})}{\int d^{2} 
{\bf b}\ G_{in} (b,s)}
= \Phi(Z,s),
\end{equation}
where

\begin{equation}
\xi=\frac{\int db^{2}G_{in}(b,s)}{\int db^{2}\ G_{in}(b,s) 
\Omega^{\gamma}(b,s)} = \xi(s).
\end{equation}

We proceed with the determination of the elementary distribution 
$\varphi$ 
and the power coefficient $\gamma$, through quantitative analyses 
of 
$e^{+}e^{-}$ data and under the following arguments. Because the 
elementary 
process occurs at a given impact parameter, its elementary structure 
suggests 
that it should scale in the KNO sense. Now, since experimental 
information 
on $e^{+}e^{-}$ multiplicity distributions shows agreement with this 
scaling, we shall base our parametrization for $\varphi$ just on 
these data. 
In particular, it is sufficient to assume a gamma distribution, 
normalized 
according to Eq. (2), 
$
\varphi(z)=2 K^{K} z^{K-1} exp\{-Kz\}/ \Gamma (K) .
$
Fit to the most recent data, covering the interval 22.0 $GeV$ 
$\leq$ $\sqrt{s}$ $\leq$ 161 GeV furnished $K=10.775$ $\pm$ 
0.064 with $\chi^{2}/N_{DF}$=508/195.

Finally, we consider fits to the $e^{+}e^{-}$ average 
multiplicity through a general power law
$<n>_{e^+ e^-}= \alpha (\sqrt{s})^{\gamma}$.
We collected the most recent experimental data at the highest 
energies, 
covering the interval 5.1 GeV $\leq$ $\sqrt{s}$ $\leq$ 183 GeV. 
Fitting to this equation yields $\alpha = 2.06 \pm 0.02$ and
$\gamma=0.522 \pm 0.002$,
with $\chi^{2}/N_{DF}$=354/45.
The above parametrization deviates from the data above 
$\sqrt{s}$ $\sim$ 100 GeV and this contributes to the high 
$\chi^{2}$ 
value. However, as commented before, we do not expect that 
$e^{+}e^{-}$ 
annihilation exactly represent the collisions between hadrons 
constituents.

\section{Results and Conclusions}

With the above results we are now able to predict the hadronic 
multiplicity distribution $\Phi(Z,s)$, Eq. (6), without free 
parameters: $G_{in}(b,s)$ (and $\Omega(b,s)$) comes from analysis 
of the elastic scattering data and $\varphi(z)$ 
and $\gamma$ from fits to $e^{+}e^{-}$ data. 
We express $\Phi$ in terms of the scaling variable 
$Z^{'}=N^{'}/<N^{'}>$ where $<N^{'}>=N(s)-N_{0}$ with $N_{0} = 0.9$ 
leading charges removed. 
The predictions for $pp$ scattering at 52.6 GeV and $\overline{p}p$ 
at 546 GeV are shown in Figs. 1 and 2, together with 
the experimental data.
The behaviour of the normalization factor $\xi(s)$, as determined 
from
Eq. (7), is shown in Fig. (3) in a wide interval of energy. The
multiplicity functions $m(b,s)$, Eqs. (1) and (5), corresponding to 
the energy and
reactions of Figs. 1 and 2 are displayed in Fig. 4.

From Figs. 1 and 2, we conclude that, in the context of the 
geometrical 
approach and based 
on the BEL behaviour for $G_{in}(b,s)$, our guess for the 
elementary 
hadronic quantities (based on geometrical arguments and 
quantitative 
analysis of $e^{+}e^{-}$ data) yields excellent predictions 
for the 
hadronic distribution $\Phi(Z,s)$.

The fact that the average multiplicity associated with hadron 
constituents,
increases faster then $e^+ e^-$ data above $\sim 100$ GeV
 seems quite reasonable, since at these energies we expect 
additional 
contributions from gluons/quarks interactions which are not 
present in 
lepton-lepton collisions. 

Anyway, despite the success of the geometrical approach in the
phenomenological context, it is in general difficult to obtain
direct connections with a microscopic theory. Some aspects of these
 connections have been recently discussed \cite{bmv} and other
results, along these lines, were also presented at this Conference
\cite{geo}.

\section*{Acknowledgments}
M.J.M. is thankful to Professor V. A. Petrov and the whole Organizing 
Committee
of the Conference for the kind hospitality and the excellente atmosphere 
of the
meeting.
Thanks are also due to FAPESP and CNPq for 
financial support.

\twocolumn

\begin{figure}[t]
\centerline{
\psfig{figure=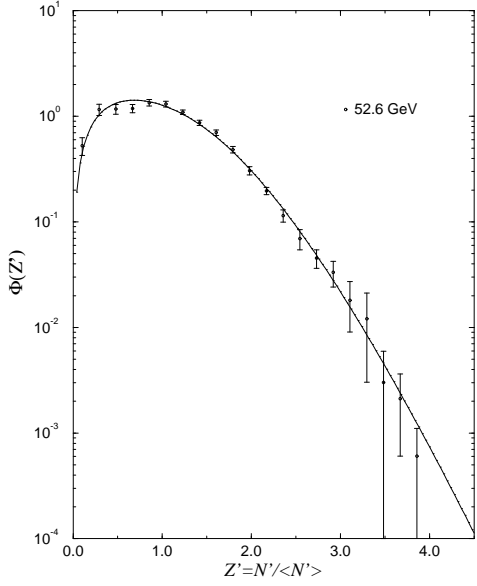,height=2.7in}
}
\vskip0.9truecm
\caption{Scaled multiplicity distribution for inelastic $pp$ data
at ISR energies compared to theoretical expectations using Eq. (6).
\label{fig1}}
\end{figure}

\begin{figure}[t]
\centerline{
\psfig{figure=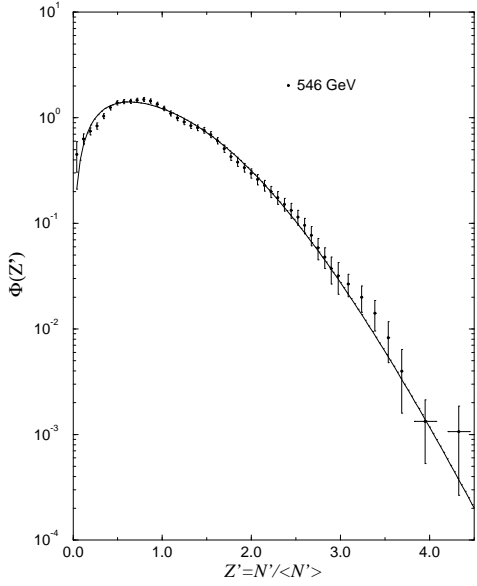,height=2.7in}
}
\vskip0.9truecm
\caption{Same as Fig. 1 for inelastic $\overline{p}p$ data
at collider energies. \label{fig2}}
\end{figure}

\begin{figure}[t]
\centerline{
\psfig{figure=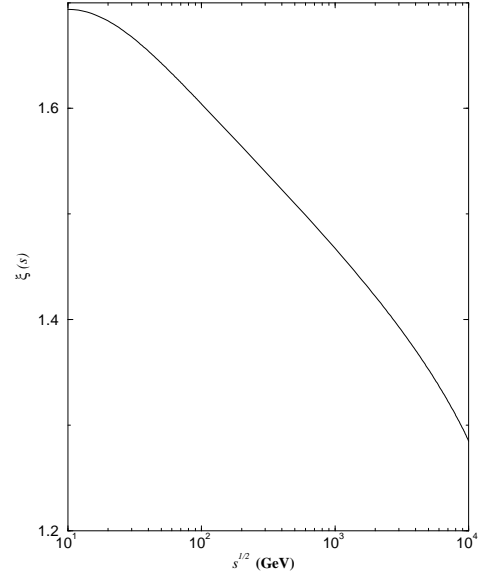,height=2.7in}
}
\vskip0.9truecm
\caption{Normalization factor from Eq. (7) as function of the energy. 
\label{fig3}}
\end{figure}

\begin{figure}[t]
\centerline{
\psfig{figure=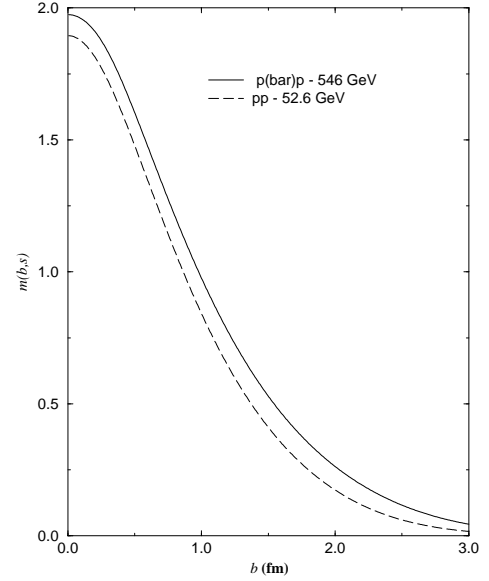,height=2.7in}
}
\vskip0.9truecm
\caption{Multiplicity function from Eq. (5) for $pp$ at 52.6 GeV and
$\overline{p}p$ at 546 GeV. \label{fig4}}
\end{figure}

\end{document}